# Room temperature giant baroresistance and magnetoresistance and its tunability in Pd doped FeRh

Pallavi Kushwaha*, Pallab Bag, and R Rawat†

*UGC-DAE Consortium for Scientific Research, University Campus, Khandwa Road, Indore-452 001, India*

We report room temperature giant baro-resistance ($\approx$128%) in $Fe_{49}(Rh_{0.93}Pd_{0.07})_{51}$. With the application of external pressure and magnetic field the temperature range of giant baro-resistance ($\approx$600% at 5K and 19.9 kbar and 8 Tesla) and magnetoresistance ($\approx$-85% at 5K and 8 tesla) can be tuned from 5 K to well above room temperature. As the AFM state is stabilized at room temperature under external pressure, it shows giant room temperature magnetoresistance ($\approx$-55%) with magnetic field. Due to coupled magnetic and lattice changes, the isothermal change in room temperature resistivity with pressure (in the absence of applied magnetic field) as well as magnetic field (under various constant pressure) can be scaled together to a single curve when plotted as a function of X = T + 12.8*H - 7.2*P.

First order antiferromagnetic (AFM)-ferromagnetic (FM) transition in equiatomic FeRh and its derivative has been of interest for giant magnetoresistance (MR),[1–3] magnetostriction,[4–7] magnetocaloric effect (MCE),[8–10] glass like magnetic states[3,11] etc. The origin of AFM-FM transition, which is accompanied with large isostructural unit cell volume change and change in electrical resistance,[12] is still a subject matter of theoretical investigation. However, extensive experimental studies on this system has provided some understanding and empirical rules to tailor its functional properties. Recent demonstration of room temperature antiferromagnetic memory resistor[13] and electric field control of magnetic order[14] are examples, where these functional properties has been utilized. On the other hand, correlation between 'e/a' ratio and first order transition temperature ($T_N$) has been shown by Barua et al.[15] which in turn is used for synthesizing new alloys with Cu and Au substitution. Similarly, correlation between transition temperature and its rate of change with pressure/magnetic field has been reported for a wide variety of dopant in this system.[16–18] Study on a disorder broadened AFM-FM transition in doped FeRh system showed that pressure and magnetic field shift transition temperature but the extent of hysteresis and the width of the transition is determined by the temperature.[18] This interplay of pressure and magnetic field in FeRh system provide an opportunity to tune the critical parameters for inducing AFM-FM transition, which can be utilized for practical applications. Here, we report giant resistivity change in Pd doped FeRh with simultaneous application of pressure and magnetic field over a wide temperature range (from 5 K to more than 300 K). The value of giant resistivity change at room temperature is found to be $\approx$128% with pressure and $\approx$-55% with magnetic field, which increases to about $\approx$600% and $\approx$-85% around liquid He temperature. We show that the isothermal change in resistivity with pressure and magnetic field can be scaled together.

Polycrystalline sample used in the present study is the same as used in the earlier study.[18] Resistivity ($\rho$) and the longitudinal magnetoresistance measurements under applied pressure are performed using Cu-Be high pressure cell from easy Lab (U.K) in the temperature range of 5 – 320 K up to 21 kbar pressure and 8 Tesla magnetic field (using superconducting magnet system from Oxford Instruments). A mixture of iso-amyl alcohol and n-pentane (1:1 volume ratio) is used as a pressure transmitting medium. The pressure inside the chamber is measured using a calibrated manganin wire resistance. Pressure is varied at room temperature for all the measurements and then locked for low temperature and high field measurements. The standard four-probe method is used for measuring the sample resistance. Change in resistance is defined as $\{\rho(H/P) - \rho(0)\}/\rho(0)$, where $\rho(H/P)$ is resistivity in the presence of magnetic field (H)/pressure(P) at constant pressure (P)/magnetic field (H). Strain measurements are carried out using strain gauge (TML Tokyo Sokki Kenkyujo co. Ltd.) on flat surface of a disk ($\approx 6mm$ diameter) taken from the same ingot of the sample.

Figure 1 shows the temperature dependence of MR during cooling and subsequent warming for 8 Tesla magnetic field under labeled constant applied pressure. These MR curves are calculated from the temperature dependence of zero and 8 Tesla resistivity shown in the inset of respective figure. For these measurements magnetic field as well as pressure is applied at room temperature and resistance is measured during cooling and subsequent warming. In the absence of applied external pressure large negative MR ($\geq$-70%) below around 200 K is observed as shown in figure 1(a). The magnitude of MR reaches a value of -85% at 5 K, even though some AFM phase is retained during 8 Tesla field cooling. In spite of it these values are comparable to reported MR for Pd doped FeRh by Baranov and Barabanova et al.[12] where these are in the range $\approx$-55% to -77% depending on sample quality. Therefore observation of larger MR in the present sample indicate good quality of the sample and in fact our earlier high field isothermal MR studies (up to 14 Tesla which was sufficient enough to almost complete the transition at 5 K) showed more than $\geq$-90% MR.[3] In the presence of applied external pressure of 9 kbar, the temperature region of giant MR is limited at both low temperature and high temperature as shown in figure 1(b). The temperature region where large MR is observed has two different magnetic states i.e. AFM and FM for zero and 8 Tesla field cooling, respectively. At lower temperature/high temperature system tends to be in AFM/FM state for both the field condition. The low temperature decrease in MR magnitude is related to AFM-FM transformation in the presence of 8 Tesla magnetic field, whereas the high temperature decrease is associated with the AFM-FM transition in the absence of applied magnetic field. For higher values of external pressure $T_N$ is shifted to above room temperature in the absence of applied external magnetic field and it results in large MR (more than -55%) around room temperature as shown in figure 1 (c) for 14.1 kbar applied pressure.

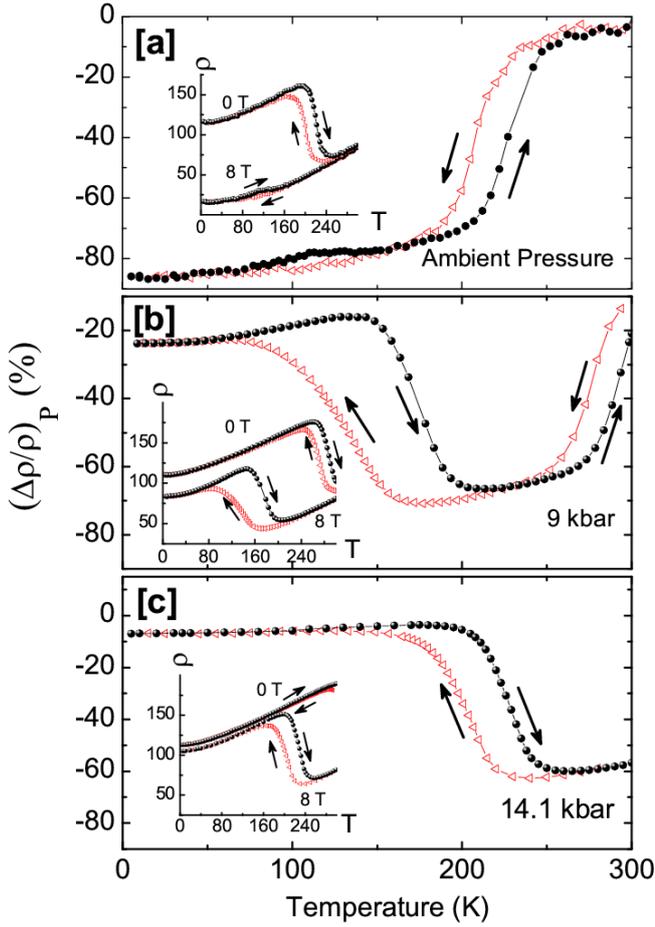

FIG. 1: (Colour online) Temperature dependence of magnetoresistace (MR) for 8 Tesla magnetic field change **[a]** in the absence of applied pressure, **[b]** in the presence of 9 kbar and **[c]** in the presence of 14.1 kbar . These MR curves are calculated from corresponding resistivity measured ($\rho$) during cooling and subsequent warming in the presence of 0 and 8 Tesla magnetic field and the labeled pressure value (applied at room temperature) and are shown as inset in the respective figure.

Figure 2 shows the change in resistance with applied pressure. For the sake of clarity curves for only warming cycle are shown. Temperature dependence of strain in the absence of applied magnetic field and pressure is shown in the inset of figure 2(a), which shows that the low temperature state has lower volume compared to high temperature FM state. As a result, pressure favors high resistance AFM state. In the absence of applied magnetic field (figure 2(a)) a large positive baroresistance with increasing pressure is observed. For lower pressure value a small peak near zero field transition is observed, whose magnitude and peak position shifts to higher temperature with increasing pressure. It can be related to broad first order AFM-FM transition which (width of the transition) is found to be about 50 K for zero field warming curve.[18] Due to this broadening, AFM and FM phase coexist over this temperature range and lower the temperature higher is the AFM phase fraction. Therefore at low T, only a small fraction of FM phase exist which can be transformed to AFM state with the application of external pressure and hence smaller magnitude of baroresistance. On the other hand at high T larger pressure is required for FM to AFM transformation. When shift in transition temperature with increasing pressure becomes larger than the transition width, a relatively flat region in baroresistance vs. temperature curve is observed as can be seen for pressure values higher than 9 kbar in the figure 2(a). The maximum value of baroresistance is found to be more than 150% around 250 K and remains well above 125% even at room temperature. In the presence of 8 Tesla magnetic field, temperature region of giant baroresistance is shifted to low temperature as shown in figure 2(b).

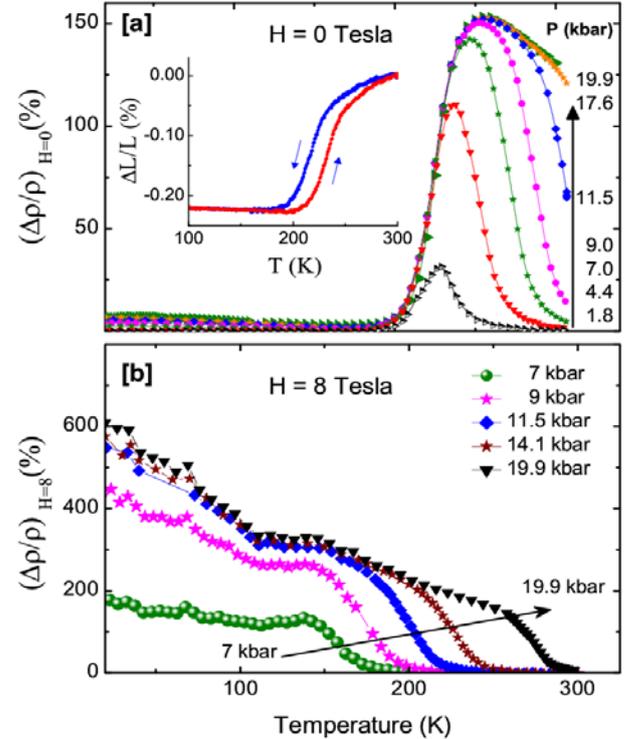

FIG. 2: (Colour online) Baro-resistance calculated from temperature dependence of resistivity **[a]** in the absence of applied magnetic field and **[b]** in the presence of 8 Tesla magnetic field for change of pressure from 0 to labeled magnetic field. For the sake of clarity only curves during warming are shown. Inset in the top figure shows relative change in sample length with temperature in the absence of applied magnetic field and pressure.

In the absence of applied pressure system remains predominantly in FM state during 8 Tesla field cooling. With increasing pressure this AFM state starts to transform to FM state at low temperature resulting in a giant baro resistance (≈600%) for pressure higher than 11.5 kbar. For lower pressure only a fraction of FM state is transformed to AFM state, which therefore give rise to smaller change in resistivity with pressure.

It has been shown earlier that magnetic state of this system depends on the path followed in HT space and therefore one can get different values of magnetoreistance calculated from resistivity measured during isothermal and field cooled cooling case.[3] It is possible that similar irreversibility could be observed for baroresistance, particularly at low temperature. However, it could not be verified experimentally with present set up as, pressure can be varied only at room temperature. Ease of applying external magnetic field compared to pressure has been one of the reason for extensive use of H in the study of magnetic glasses.[19] However, we could measure isothermal change in baroresistance at room temperature under ambient field condition. The results of this measurement are shown in figure 3. Sharp rise in resistivity with increasing pressure indicate a pressure induced transition from FM to AFM state. Critical pressure required for AFM-FM transition, taken as the pressure at which pressure derivative of resistivity show a maxima (as shown in the inset of figure3), are found to be 12.4 kabar and 10.6 kbar. The magnitude of resistivity change associated with pressure induced FM to AFM transition is ≈128% at 19.9 kbar applied pressure which is consistent with baroresistance values obtained from our isobaric resistivity measurement. It suggest absence of glass like irreversibility around 300 K in the presence of high pressure measurements.

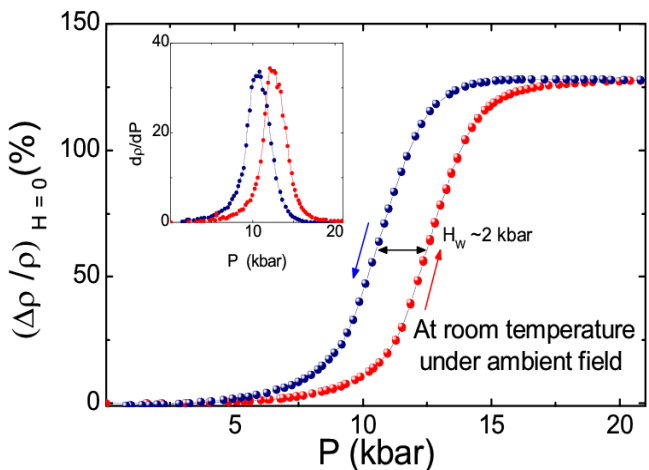

FIG. 3: (Colour online) Pressure induced FM-AFM transition at ambient temperature and magnetic field showing a giant baroresistance of ≈128%. Inset shows the pressure derivative of isothermal resistivity change.

The AFM state obtained with pressure at room temperature can be transformed back to FM state with magnetic field, which can give rise to large MR. Figure 4(a) shows the isothermal MR measured at 300 K under various constant applied pressure values for which system is in AFM state in the absence of applied magnetic field. The magnitude of MR associated with field induced AFM to FM transition is found to be ≈55% and the critical field required for AFM to FM transition (summarized in the table 1) increases almost linearly with pressure at a rate of 0.58 Tesla/kbar around 300 K. The magnitude of room temperature MR is higher than that observed for FeRh (around 50% for ≥12 Tesla) by Algarabel et al.[1] The reason for higher MR could be better quality of the sample or change in electronic structure with Pd substitution or both. Manekar et al.[10] also speculated the change in electronic structure for enhanced Magnetocaloric effect in their Ni doped FeRh sample when compared to FeRh. In fact change in valence electron per atom for Ni as well as Pd substitution and its effect on first order transition is shown to be similar.[15]

TABLE I: Critical field in different applied pressure at room temperature.

| Applied pressure | $H_{up}$ | $H_{dn}$ |
|---|---|---|
| 19.9 kbar | 5.799 T | 4.739 T |
| 17.6 kbar | 4.596 T | 3.425 T |
| 14.1 kbar | 2.51 T | 1.268 T |

Opposing influence of P and H on $T_N$ has been used to study the role of P, H and T in determining the width of transition and the extent of hysteresis[18]. This study also showed that field derivative of isothermal magnetoresistance measured for various P values for a given T have similar field dependence except for shift in critical field. As shown in figure 4(b), this equivalence can be extended for not only to field dependence but also for pressure dependence of resistivity change. Here the relative change in resistivity (with respect to resistivity of AFM state i.e. $\rho_{AF}$) for all the four curve shown in figure 3 as well as in figure4(a) are plotted as a function of X, which is defined as X= T + 12.8*H - 7.2*P (where T is the temperature of measurement i.e. ≈ 305 K for isofield (figure 3) and 300 K for isobaric measurement). The coefficient of second and third term in this equation are related with the rate of change of $T_N$ around room temperature with magnetic field and pressure, respectively and these are taken as constant to a first approximation. This scaling shows that magnetic field and pressure can be treated equivalently i.e. magnetic field effectively acts as a negative pressure (decompression) and vice-versa. This equivalence between P and H could be useful from application point of view, where large effect are required for smaller change in applied perturbation. Therefore by suitable choice of P and H larger change in resistivity can be achieved for much smaller change in P and H, when applied simultaneously.

To conclude, we studied MR and baroresistance associated with FM to AFM transition in $Fe_{49}(Rh_{0.93}Pd_{0.07})_{51}$ and showed the tunability of magnetoresistance and baroresistance over a wide temperature range. Isothermal application of pressure at room temperature induces FM to AFM transition resulting in more than 125% change in resistivity. Application of magnetic field transform the pressure induced AFM state back to FM state resulting in a room temperature giant negative MR −55%. The critical fields required for transition between FM and AFM state varies linearly with pressure at room temperature. For practical applications one need large changes in resistivity for smaller change in pressure or magnetic field values. These results indicate that it can be achieved over a wide temperature range with simultaneous application of pressure and magnetic field in suitably doped FeRh system. Further, earlier studies have shown correlation between MR and MCE[20–22] and therefore MR can be used to predict MCE behaviour. The equivalence between pressure and magnetic field demonstrated in this study will also be useful in predicting MCE and elastocaloric properties from resistivity studies under magnetic field or pressure.

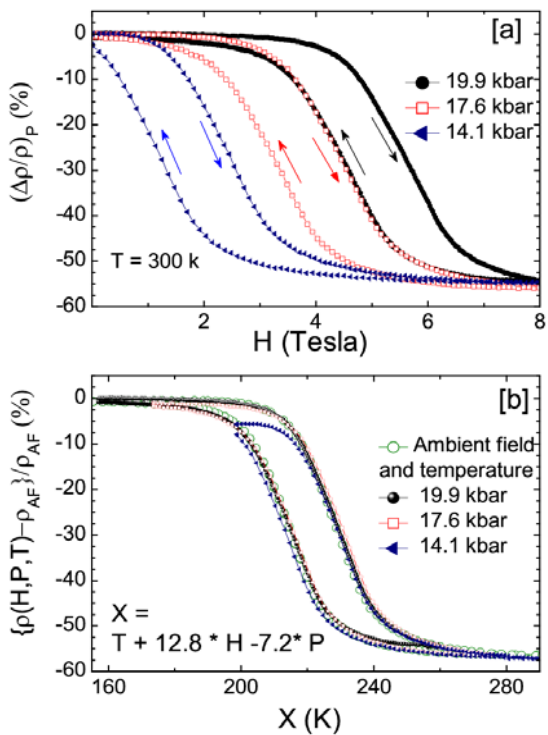

FIG. 4: (Colour online) **(a)** Critical field tunebility for back transformation of pressure induced AFM state to FM state giving rise to giant room temperature magnetoresistance **(c)** MR curve shown in figure (a) as well as baroresistance shown in figure 3 can be scaled to a single curve where X = T + 12.8*H - 7.3 * P and y axis represent % change in resistivity for FM to AFM transition. See text for details.

## Acknowledgments

Sachin Kumar is acknowledged for help during resistivity measurements. PK acknowledges CSIR, India for Senior Research Fellowship.